\documentstyle[aaspp4,epsf,11pt]{article}

\newcommand{\rosat}{ROSAT}
\newcommand{\asca}{ASCA}
\newcommand{\psec}{s$^{-1}$}
\newcommand{\pyr}{yr$^{-1}$}
\newcommand{\parcmin}{arcmin$^{-2}$}
\newcommand{\df}{\dotfill}

\def\pcm#1{cm$^{-#1}$}                

\newcommand{\nsnr}{n_{\rm SNR}}
\newcommand{\moo}{M\"{O}}
\newcommand{\etal}{{et al.}}
\newcommand{\rhr}{$^{\rm h}$}
\newcommand{\rmin}{$^{\rm m}$}
\newcommand{\rsec}{$^{\rm s}$}
\newcommand{\EM}{{\rm EM}}

\slugcomment{Accepted for publication in {\it The Astrophysical Journal, Letters}.}

\begin{document}

\title{The \asca\ Spectrum of the Vela Pulsar Jet}
\author{Craig B. Markwardt and Hakk\i\ B. \"Ogelman}
\affil{University of Wisconsin --- Madison,
       Department of Physics,
       1150 University Avenue,
       Madison, Wisconsin 53706}

\begin{abstract}
\rosat\ observations of the Vela pulsar and its surroundings revealed
a collimated X-ray feature almost 45\arcmin\ in length (Markwardt \&
\"Ogelman 1995), interpreted as the signature ``cocoon'' of a
one-sided jet from the Vela pulsar.  We report on a new \asca\
observation of the Vela pulsar jet at its head, the point where the
jet is believed to interact with the supernova remnant.  The head is
clearly detected, and its X-ray spectrum is remarkably similar to the
surrounding supernova remnant spectrum, extending to X-ray energies of
at least 7~keV.  A \rosat+\asca\ spectrum can be fit by two-component
emission models but not standard one-component models.  The lower
energy component is thermal and has a temperature of $0.29\pm0.03$~keV
($1\sigma$); the higher energy component can be fit by either a thermal
component of temperature $\sim$4~keV or a power law with photon index
$\sim$2.0.  Compared to the \rosat-only results, the mechanical
properties of the jet and its cocoon do not change much.  If the
observed spectrum is that of a hot jet cocoon, then the speed of the
jet is at least 800~km~\psec, depending on the angle of inclination.
The mechanical power driving the jet is $\ge 10^{36}$~erg~\psec, and
the mass flow rate at the head is $\ge 10^{-6} M_{\sun}$~\pyr.  We
conclude that the jet must be entraining material all along its length
in order to generate such a large mass flow rate.  We also explore the
possibility that the cocoon emission is synchrotron radiation instead
of thermal.

\end{abstract}

\section{Introduction}

While isolated pulsars are known to be losing rotational energy over
time, it is unclear precisely how this energy escapes from the pulsar,
since direct radiation accounts for only a small fraction.  Most
theories hold that the balance of the power leaves the pulsar in the
form of a coherent particle outflow.  The emission from the Crab
Nebula, for example, can be explained as synchrotron radiation from
high-energy electrons which have been accelerated in a shocked outflow
from the pulsar.  The luminosity of the entire nebula is consistent
with the spindown luminosity of the Crab pulsar.  The Vela pulsar,
PSR~0833--45, has a spindown age of $\tau\simeq 10^4$~yr, an estimated
distance of 500~pc, and a rotational energy loss rate of $7\times
10^{36}$~erg~\psec (Taylor, Manchester \& Lyne 1993).  A $\sim$0.5~pc
diameter high energy compact nebula surrounds the pulsar (de Jager,
Harding \& Strickman 1996; \"Ogelman, Finley \& Zimmerman 1993;
Harnden, \etal\ 1985) but its X-ray luminosity is only $\sim
10^{33}$~erg~\psec, too small to explain the pulsar energy loss.
Recent X-ray observations by \rosat\ (Markwardt \& \"Ogelman 1995,
hereafter \moo), suggest that there is a polar outflow which forms a
jet interacting with the supernova remnant (SNR; see
Figure~\ref{rosat_im}).  They interpret the observed X-ray structure
to be thermal emission from a shock-heated cocoon of gas surrounding
the jet itself, and conclude that the mechanical energy required to
drive the jet into the SNR interior is comparable to the spindown
luminosity of the Vela pulsar.  Spatially, the jet is $45\arcmin\times
12\arcmin$ and centrally brightened, corresponding to
$6.5$~pc~$\times$~1.7~pc at a distance of 500~pc, and coincides
roughly with the radio feature Vela~X.  \moo\ determined that the
cocoon spectrum is spatially uniform, described by a thermal plasma
model with a temperature of 1.3~keV and density 0.4~\pcm{3}, apparent
only above 0.7~keV.  Below that energy, emission from the Vela SNR as
a whole dominates the spectrum.  The spectral fit to the \rosat\ data
is not unique however: a power-law is also satisfactory but
overpredicts the jet emission when extended to lower energy X-rays.

This paper describes an \asca\ observation of the ``head'' of the Vela
jet, the point where the the jet is believed to interact with the SNR.
We find that the spectrum is remarkably smooth up to 7~keV, with
perhaps only one strong emission line near 1~keV. In sections 2 and 3,
we describe the observations by ASCA, and present the combined
\rosat+\asca\ spectrum of the cocoon.  In sections 4 and 5, we compare
the spectral results against candidate jet models, including the
cocoon model of \moo\ and a synchrotron emission model.

\section{Observations and Analysis}

\begin{figure}
     \framebox{\epsfxsize=\textwidth\epsfbox{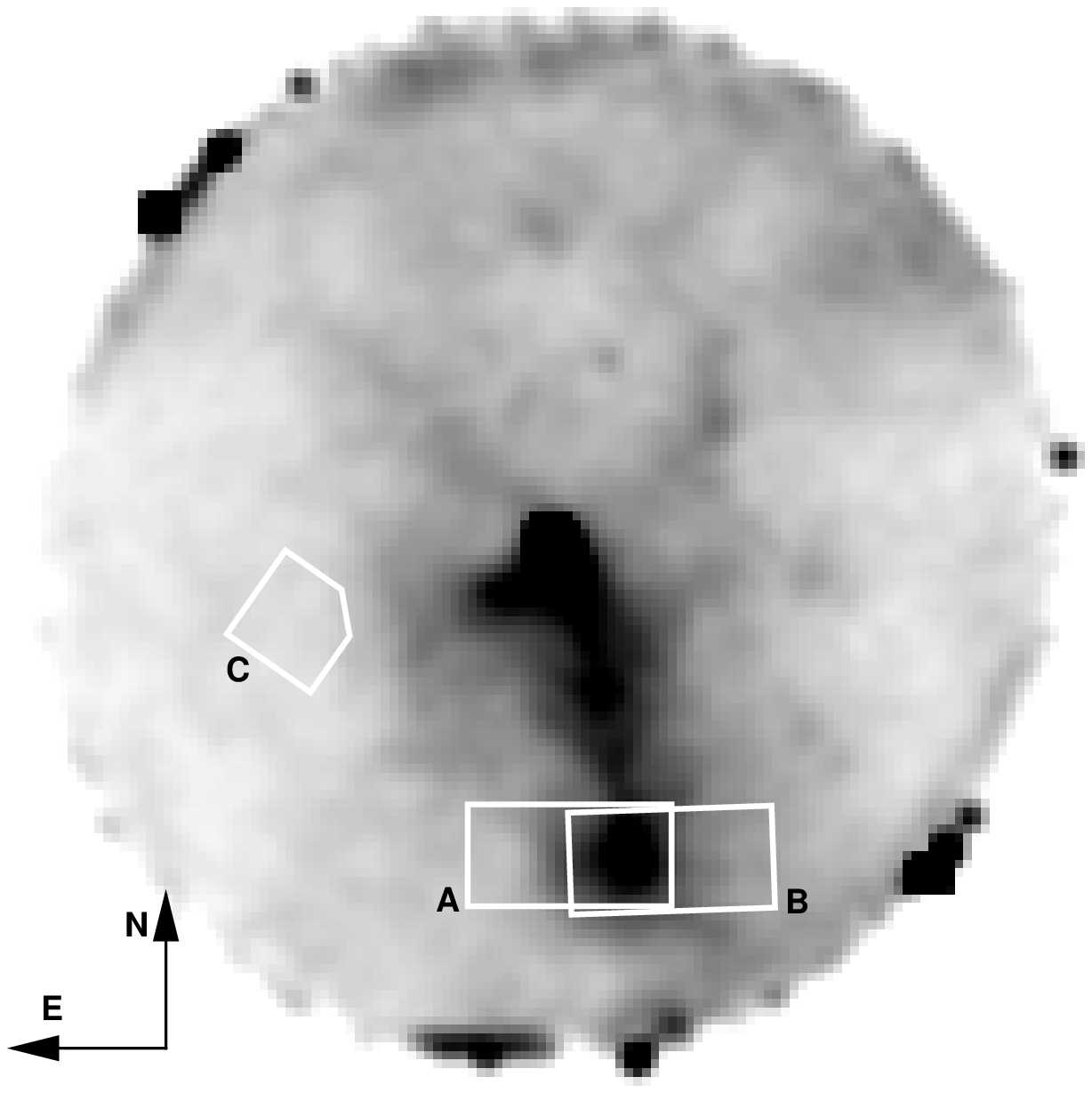}}
\caption{Image of the Vela pulsar and its 1\arcdeg-radius surrounding
field, as obtained by \rosat\ in the 0.7--2.4~keV X-ray band.  The jet
projects from the pulsar (at center) to south-southwest
$\sim$45\arcmin. The \asca\ fields used in this work are indicated as
white boxes: (A) \& (B) are observations of the jet; (C) is an
archival observation (sequence 50021010) of the Vela SNR used for
additional background measurements.  The centers of fields (A) and (B)
were, respectively, $\alpha(2000)$=08\rhr34\rmin55\rsec,
$\delta(2000)$=$-$45\arcdeg52\arcmin13\arcsec, and
$\alpha(2000)$=08\rhr33\rmin56\rsec,
$\delta(2000)$=--45\arcdeg52\arcmin13\arcsec.\label{rosat_im}}
\end{figure}

\asca\ observed the ``head'' of the Vela jet on 1995~April~15--16, in
two separate pointings of 15~ks usable time each.  SIS observations
were conducted in 2-CCD Bright mode with the long chip axis oriented
nearly perpendicular to the jet axis.  Figure~\ref{rosat_im} shows the
\rosat\ image of the jet in the X-ray band 0.7--2.4~keV, with a
representation of the position and orientation of the two \asca\
pointings (labelled pointings A \& B).  The 2-CCD configuration has an
11\arcmin$\times$22\arcmin\ field of view, but the pointings partially
overlapped for the purpose of sampling the SNR emission on either side
of the jet cocoon, leaving a net field of view of
11\arcmin$\times$35\arcmin.  For both exposures, the cocoon head was
primarily in the field of view of one CCD, leaving the other CCD to
sample the SNR emission.  \asca\ is an imaging telescope, but for the
purposes of this {\it Letter\/}, we have only used the spatial
resolution of the instrument to distinguish between emission from the
cocoon and from the patches of SNR immediately adjacent to the cocoon
on the sky --- which we consider to be ``background.''  \rosat\ and
\asca\ spectra were obtained using identical selection regions.
\asca\ X-ray events were screened using the criteria recommended by
the ASCA Guest Observer Facility to remove periods of high particle
background.  Analysis was performed with the FTOOLS package,
version~3.3, and with XSPEC version 9.00.  Selection regions did not
cross the CCD boundaries.  The region including the head of the cocoon
had a count rate of (5.3--5.9)$\times 10^{-3}$~ct~\psec~\parcmin\
compared to the surrounding SNR count rate of (2.6--3.8)$\times
10^{-3}$~ct~\psec~\parcmin\ (per CCD), where the variation is due to
different CCD sensitivities.

Because the Vela SNR itself contributes about half of the observed
\asca\ counts from the cocoon, background subtraction is a concern.
Background regions were taken from the outermost
11\arcmin$\times$6\arcmin\ ``wings'' of the field of view where cocoon
emission was the lowest.  We were concerned that the jet emission
could contaminate the background spectrum because of the broad PSF of
the telescope.  The background we observe is very similar to SNR
emission seen in an archival \asca\ observation of a portion of the
SNR at least 30\arcmin\ from the jet (shown in Figure~\ref{rosat_im}),
leading us to conclude that the broad PSF of \asca\ is not a
signifigant contaminating factor.  We also deem it unlikely that the
effect of stray single-reflection X-ray photons from off-axis sources
(including the Vela SNR itself) makes a large contribution to the
cocoon spectrum, since the photons would need to be preferentially
imaged onto the cocoon region in both SIS exposures of the
observation.  Because all of the \rosat\ and \asca\ spectra are self
consistent where they overlap in energy coverage, we infer that any
\asca-specific contamination effects are small.

\section{Spectrum}

\begin{figure}
    \framebox{\epsfxsize=\textwidth\epsfbox{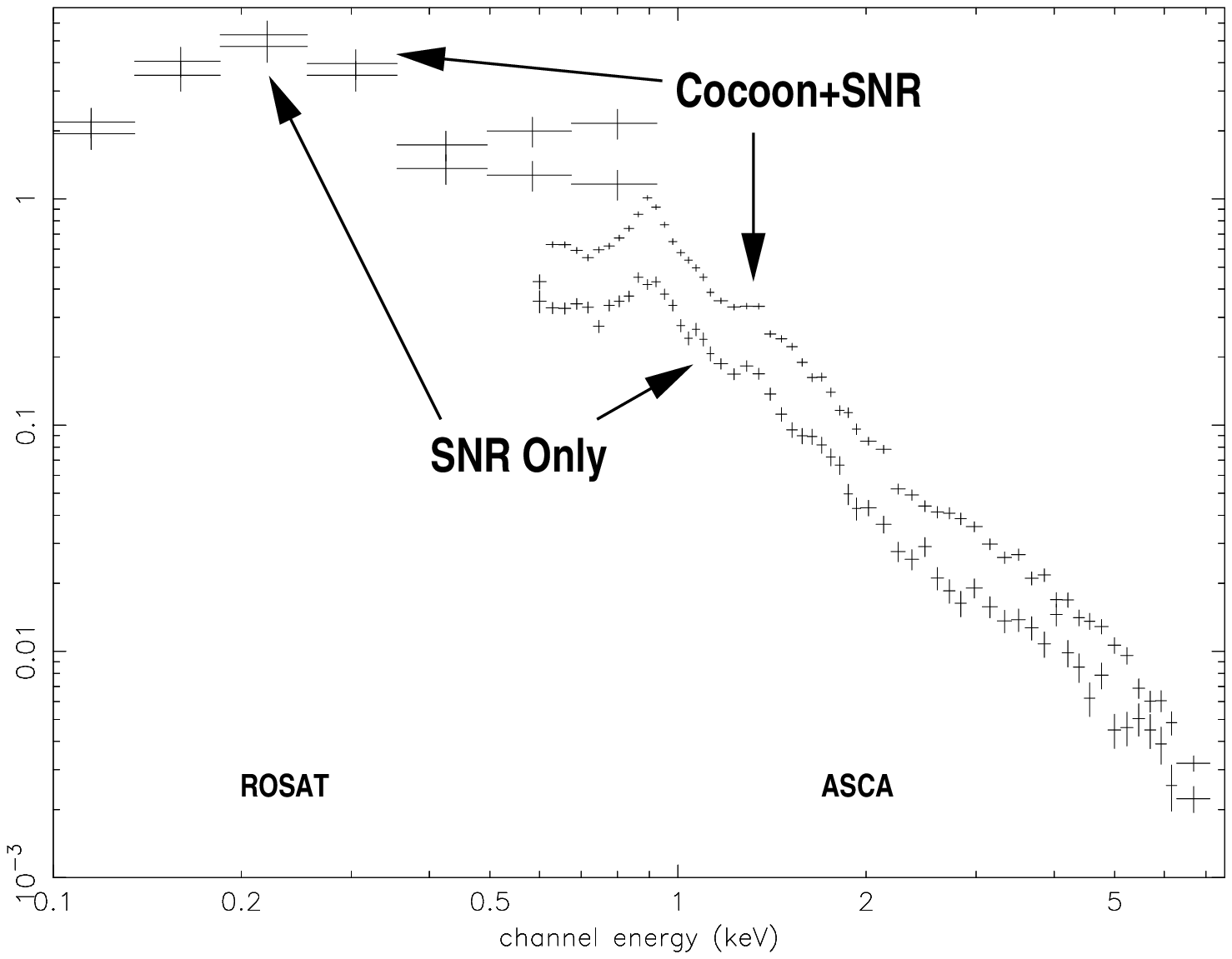}}
\caption{Combined \rosat\ and \asca\ spectra of the Vela jet$+$SNR
{\it vs.\/} SNR alone, in units of cts~\psec~keV$^{-1}$ per SIS CCD
solid angle of $11\times 11$~arcmin$^2$.  The \rosat\ spectra range
from 0.1--0.9~keV, and are represented by the wide data bins; the
\asca\ spectra range from 0.6--7.0~keV and are represented by narrow
bins.  The jet is the more intense (upper) spectrum, as indicated by
the arrows.  The \rosat\ and \asca\ count rate spectra do not match
between 0.6 and 0.9~keV because of differing instrumental response.
\label{asca_spec}}
\end{figure}

Figure~\ref{asca_spec} shows the spectrum of the SNR (lower) and
cocoon$+$SNR (upper) as measured by both \asca\ and \rosat.  The
spectra are virtually identical from 0.7--7~keV except in intensity;
below 0.7 keV the SNR emission dominates the spectrum.  Various
spectral models were fitted to the combined \rosat/\asca\ spectrum of
the cocoon.  Single-component models produced very poor fits and are
not shown.  The best fits had two components: a lower temperature
thermal plasma ($T_l = 0.29\pm0.03$~keV $\simeq (3.4\pm 0.3)\times
10^6$~K), and a higher energy component, which is fit by either a
higher temperature plasma or power law emission.  The best-fitting
two-component models are presented in Table~\ref{modeltab}.  The
spectrum of the surrounding SNR was fit to a similar model, with the
addition of a {\it third\/} thermal component, of temperature 0.14~keV
and emission measure 0.46~\pcm{6}~pc, to represent the contribution in
the 0.1--0.7~keV range which dominates in the \rosat\ soft channels.
Because the third, cool, component was confined to only the lowest
energy channels it had little effect on the other two hotter
components.  We believe that the lowest-temperature emission is from
the outermost shells of the SNR and thus unimportant in the analysis
of the jet, as explained in more detail below.  The total unabsorbed
flux from the cocoon is 1.3--1.6$\times
10^{-13}$~erg~\pcm{2}~\psec~\parcmin\ in the 0.1--7.0~keV band for
both components, independent of the model taken, which corresponds to
a luminosity of $\sim 2\times 10^{33}$~erg~\psec\ for a
45\arcmin$\times$12\arcmin\ jet at 500~pc.

\placetable{modeltab}

Thermal models represent emission from an optically thin plasma at
collisional equilibrium (Raymond \& Smith 1977, RS), and did not
change appreciably when alternate plasma models were used (Mewe,
Kaastra \& Liedahl 1995).  The best-fit abundances are quoted in
Table~\ref{modeltab}, but we cannot be certain enough about the
detailed accuracy of the spectral models (especially at energies of
$\sim$1~keV) to claim that the quoted values represent true physical
abundances in the plasma.

\section{Cocoon Model}

We now wish to compare the spectral results to the cocoon model of
\moo.  In the cocoon model, the pulsar produces a thin collimated
outflow or jet beam, at a speed $v_j$, which interacts with the SNR at
the working surface, travelling at a speed $v_h$.  Shock heating at
the working surface causes the gas to expand laterally into a cocoon
of hot gas surrounding the jet beam.  A large fraction of the Vela
pulsar spindown luminosity drives the outflow, which in turn goes into
heating the gas and expanding the cocoon.  At the observed density and
temperature the gas has a much longer radiative lifetime ($\sim
10^7$~yr) than the pulsar's spindown age ($\sim 10^4$~yr) and so is
not in a radiation-dominated phase.  The observed X-ray emission is
thus from the hot cocoon plasma and not the jet itself.

The cocoon model of \moo\ requires that we know the cocoon and SNR
density ratio.  We estimate the electron number density from
$n=\sqrt{\EM/fl}$, where EM is the observed emission measure, $l$ is
the line-of-sight thickness of the emitting region (which we assume to
be equal to the transverse dimension as measured by the angular
diameter), and $f$ is the fraction of the volume filled by gas.
Density obtained in this manner scales as $d^{-0.5} f^{-0.5}$, where
$d$ is distance.

We expect that the center of the Vela SNR is filled by the hottest
plasma.  Thus the hottest SNR spectral component should represent the
gas at the center of the Vela SNR which surrounds the jet cocoon.  The
lower temperature components of the observed \asca/\rosat\ SNR
spectrum would be emission from the outer, cooler shells of the SNR
which are also subtended by the \rosat\ and \asca\ lines of sight.  If
we further assume that the remnant is in pressure equilibrium, then
the observed emission measures indicate that the thickness of the cool
shell is a factor $\sim 100$ times thinner than the hot centermost
gas, and thus can be ignored in applying the cocoon model.  The
assumption of pressure equilibrium is proper as long as the gas has
not begun to undergo significant radiative cooling, which as
previously mentioned is the case.  At the assumed SNR distance of
500~pc, the central electron density is $\nsnr = 0.056\pm
0.003$~\pcm{3}, where we have used a SNR angular diameter of
$7.2^\circ$ (Aschenbach, Egger \& Tr\"umper 1995).

The parameters of the cocoon depend to some extent on the angle of
inclination, $\theta$, of the jet axis to our line of sight.  The
length of the cocoon, for example, is 6.5~pc$/\sin\theta$ ($\propto
d$).  The average bow shock speed, $v_h$, estimated as the cocoon
length divided by the pulsar age, $\tau$, is
$v_h=570$~km~\psec$/\sin\theta$ ($\propto d \tau^{-1}$).  The electron
density of the cocoon --- accounting for the effect of projection on
the emission measure --- is
$n_c=0.32\pm0.02$~\pcm{3}~$\sqrt{\sin\theta}$ based on the emission
measure of the hottest component.  Since the sound speed of an ideal
monoatomic gas at a temperature of $\sim$4~keV is $\sim$700~km~\psec,
the bow shock can be supersonic in the SNR only if the jet is inclined
to our line of sight with $\sin\theta \lesssim 0.77$.  If we also
assume that the jet terminates inside the SNR radius, we arrive at the
approximate constraint, $10^\circ\lesssim\theta\lesssim 60^\circ$.

The mechanical properties of the jet can be found by considering its
interaction with the surrounding remnant at the working surface.
Proceeding in the same manner as \moo, we obtain a relation for the
relative jet speed as it reaches the head --- $v_j/v_h$ --- in terms
of the density contrast between the cocoon and the SNR.  We find $1.4
< v_j/v_h < 2.3$, or in absolute terms $v_j=$800--1300~km~\psec
$/\sin\theta$ ($\propto d\tau^{-1}$).  The mechanical power delivered
to the working surface by the jet is $\dot{E}_j = \case{1}{2}A_j
\rho_j v_j^3 = 10^{36} - 10^{37}$~erg~\psec $/\sin\theta$ ($\propto
d^{4.5}\tau^{-3}$), which is consistent with the Vela pulsar spindown
luminosity of $7\times 10^{36}$~erg~\psec.  Since a majority of the
spindown luminosity goes to heating the cocoon gas in this model, we
can also estimate the average rate of energy input via heating as the
time rate of change of the internal energy, or $\dot{E}_{th} =
\case{3}{2}(2n_c)kT_cV_c/\tau \simeq 10^{36}$~erg~\psec
$/\sqrt{\sin\theta}$ ($\propto d^{2.5}\tau^{-1}$), where $V_c$ is the
volume of the cocoon.  Such an estimate is again appropriate because
the gas cooling timescales are much longer than the SNR's present age:
all energy added to the cocoon by the pulsar is still effectively
trapped there.  Arriving at the same order of magnitude for the jet
power by two different techniques which have different distance and
age dependences, we are confident that the jet and cocoon of this
model are pulsar driven.

There are some problems with the cocoon interpretation, primarily that
the mass flow rate delivered by the jet beam to the working surface,
$\dot{M}_j = A_j\rho_j v_j = (1-10)\times 10^{-6} M_{\sun}$~\pyr
$/\sin\theta$ ($\propto d^{2.5}\tau^{-1}$), is quite large.  There is
no way for so much mass to be extracted from the surface of a
$1M_{\sun}$ neutron star with an energy budget of $7\times
10^{36}$~erg~\psec.  It is also more than the pulsar could have swept
up via its proper motion of 100~km~\psec\ to the northwest (\"Ogelman,
Koch-Miramond, Auri\'ere 1989; Bailes, \etal\ 1990) through a plasma
with density $\nsnr$.  Additional material may be entrained along
the length of the jet, converting the jet from a presumably
relativistic outflow originating at the pulsar, to a slow heavy
massive jet when it reaches the working surface.  Considering that the
jet axis is nearly perpendicular to the pulsar proper motion direction
the jet may indeed be sweeping up SNR material all along its length.
Because the derived jet speed is barely supersonic and the
temperatures of the cocoon and SNR are indistinguishable, within
errors, this seems to be the most favorable interpretation of the
cocoon model.

The cocoon also appears to be too highly collimated for a canonical
jet interaction.  Typical jet cocoons expand in the transverse
direction because they are overpressured, and are narrower at the head
than at the power source, since the oldest portion of the cocoon is
nearest the source and thus has had the most time to expand.  The Vela
cocoon as measured in X-rays has a pressure of $\sim 4\times
10^{-9}$~erg~\pcm{3} --- much larger than the surrounding SNR pressure
of $\sim 1\times 10^{-9}$~erg~\pcm{3} --- and yet appears to be
cylindrical with no widening near the pulsar.  Although magnetic
fields are known to exist and be aligned along the jet axis (Milne
1980, 1995), the radio observations of Frail, \etal\ (1996) do not
find enough magnetic pressure to confine the jet transversely either.

\section{Synchrotron Model}

The power law model fits equally well to the \rosat+\asca\ spectrum.
If the cocoon spectrum is a power law, then we may be observing
synchrotron radiation from a relativistic plasma in the presence of a
magnetic field.  The existence of magnetic fields in the region is
well established, and radio emission from the outskirts of the cocoon
region implies that some form of particle acceleration mechanism is
indeed occurring (Frail, \etal\ 1996).  Furthermore, previous
non-imaging hard X-ray spectra of the Vela~X region are power laws,
with photon indices ranging between 2.1--2.3 (Smith \& Zimmermann
1985; Pravdo, \etal\ 1978), and are compatible with our \asca\ spectra
of both the cocoon and the SNR.  Finally, when we examine the \rosat\
data for the full $1^\circ$ field surrounding the pulsar, we find a
large region covering 70--80\% of the field which has a spectrum
similar to our \asca\ spectrum (at least, to the sensitivity level
possible with \rosat).  These facts push us to consider the strong
possibility that the center of the Vela SNR --- and the jet in
particular --- is a site of accelerated particles radiating by the
synchrotron process.

A consistent model explaining the presence of the emission is
desirable.  A ``first draft'' of the model might be as follows.  The
pulsar produces a wind outflow, highly collimated by some mechanism,
which emerges along its spin axis.  The outflow travels outward until
it interacts with the surrounding SNR gas and is shock-accelerated.
It seems reasonable to assume that fresh particle acceleration is
occurring along the entire length of the jet, since by our estimates
the characteristic synchrotron lifetime of the electrons producing
$\ge 1$~keV X-rays is $\le$5000~yr, less than the $10^4$~yr age of the
SNR.  We find it unlikely that the particles could be confined to the
jet region by any external pressure once they are accelerated.  As
mentioned previously, the exterior thermal and magnetic pressures in
the SNR are not enough to contain expansion.  We expect that the
relativistic particles will diffuse outward, perhaps to fill a large
interior portion of the SNR.  This may also explain why we observe a
large region of hard X-ray emitting plasma with \rosat\ and the
presence of a power law in the \asca\ SNR spectrum.  The jet region
has the highest density of accelerated particles (since it is the
source), and thus radiates with the highest intensity; particles which
diffuse into the lower-density surroundings also emit synchrotron
radiation, but at a lower intensity.

\section{Conclusion}

We have detected emission from the Vela pulsar jet cocoon with the
\asca\ X-ray telescope.  It is at least partially thermal with a peak
near 1~keV, but has a smooth spectrum above that energy which extends
to at least 7~keV.  The derived cocoon-model quantities for the jet
have not changed much compared to the \rosat-only results of \moo, but
it has now become clear that the spectrum is not a single plasma or
power law emission model, but rather has at least two components.

There are some other uncertainties which need to be resolved.  The
cocoon model parameters quoted in this paper depend to a large extent
on accurate determinations of the pulsar distance and age.  While the
``canonical'' distance for the Vela pulsar has been long quoted as
500~pc, there have been claims by Jenkins \& Wallerstein (1995) that
this distance is not entirely compatible with the standard model of SN
energetics.  Their optical observations of blast wave filaments are
more consistent with a distance of 250~pc.  As to the Vela pulsar's
age, we have used the spindown age of $\tau = P/2\dot P$, which
assumes a braking index of $n=3$.  Lyne, \etal\ (1996) have recently
obtained a braking index of $n=1.4\pm0.2$ by examining the history of
$\dot\nu$ over a time baseline of 25~yr which includes nine major
glitch events.  If correct, this result implies that the pulsar could
be as much as two to three times older than the spindown age.  If
either the distance were smaller than 500~pc or the age of the pulsar
were more than $10^4$~yr, then the derived jet speed, $v_j$ would
become slower by the ratio $d/\tau$, which would reduce the mechanical
power of the jet in the cocoon model significantly.

The SNR and cocoon spectra are nearly identical.  Although this might
suggest that the observed jet is merely a high density filament in the
SNR, we are convinced that the jet is truly a pulsar-powered
phenomenon, on the basis of the following arguments. (1) Most of the
filaments in the Vela SNR have temperatures in the range 0.1-0.2~keV
(\moo; Kahn, \etal\ 1985), and yet the jet cocoon has at least one
component whose temperature is 3--4~keV. (2) The jet appears to emerge
along the pulsar spin axis, the only symmetry axis which is fixed in
time (Radhakrishnan \& Cooke 1969). (3) If the cocoon model is
applied, then the mechanical power of the jet is comparable to the
pulsar spindown luminosity.  (4) The thermal energy stored in the
cocoon is consistent with the pulsar spindown energy. (5) The image of
\moo\ clearly shows that the starting point for the jet is at the
pulsar's present-day position rather than its birthplace.  This last
fact also argues against the possibility that Vela's progenitor star
generated the jet structure before the supernova --- say, as a wind
interacting with the ISM.  Any such structure would be connected to
the pulsar's {\it birthplace}, roughly 8\arcmin--10\arcmin\ to the
southeast (\"Ogelman, \etal\ 1989; Bailes, \etal\ 1990) and not
today's position.

We have planned an additional observation by \asca\ of the remainder
of the jet, which is yet unobserved, in order to determine whether the
spectral properties of the jet vary along its length.  We note, for
example, the recent results of Tamura, \etal\ (1996), who find a hot
nebula with \asca\ near PSR~1509--58 in the SNR MSH~15--52.  They
observe a non-thermal X-ray feature extending linearly from the pulsar
and a hot thermal plasma ``cloud'' at the end of the jet feature,
which shows line emission by magnesium.  They interpret the
observations as a relativistic stream of particles generated by the
pulsar, radiating synchrotron emission, and upon interaction with the
ambient material at the end of the jet, also radiating thermal
emission by shock heating.  Our observations show that the Vela jet is
also radiating thermal emission from the head, in particular by neon.
At present we do not know if the thermal emission extends over the
length of the Vela jet, or --- as in MSH~15--52 --- is confined to the
head only.  Future work by \asca\ should determine this.

\acknowledgements We thank Ali Alpar, Don Cox, and Robin Shelton for
valuable conversations, and to Chris Greiveldinger and Samar Safi-Harb
for their general comments and help.  This work was supported by NASA
grants NAGW-2643 and NAG5-2557.

\newpage 

\newpage

\newpage


\begin{deluxetable}{lccc}
\tablecolumns{4}
\tablewidth{6.5in}
\tablecaption{Spectral Parameters for Vela Jet Cocoon \& SNR\label{modeltab}}
\tablehead{\colhead{Parameter} & \colhead{Cocoon} & \colhead{SNR{\tablenotemark{a}}} & \colhead{Approx.} \\
                               &                  &               & \colhead{$1\sigma$ errors}}
\startdata
$N_H$~($10^{20}$~\pcm{2})\tablenotemark{b}\df&   2.6    & 2.6  \nl
\cutinhead{1 Temperature plasma + Power law}
$kT_l$~(keV)\tablenotemark{c}            \df&  0.29 & 0.33 & $\pm0.03$\nl
EM$_l$~(cm$^{-6}$ pc)      \df&  0.15 & 0.13 & $\pm0.03$\nl
PL photon index\makebox[0.5in]{\df}
                              &  2.11 & 2.08 & $\pm0.15$\nl
PL flux\tablenotemark{d}   \df&  9.85 &10.8  & $\pm1.65$\nl
[Ne]\tablenotemark{e}      \df&  1.65 & 1.08 & $\pm0.50$\nl
[O]                        \df&  0.50 & 0.42 & $\pm0.12$\nl
[Fe]                       \df&  0.20 & 0.23 & $\pm0.08$\nl
$\chi^2$~(d.o.f.)          \df&  320.4 (239) & 570.5 (403) & \nl
\cutinhead{2 Temperature plasma}
$kT_l$~(keV)\tablenotemark{c}  \df& 0.29 & 0.34 & $\pm0.03$ \nl
EM$_l$~(cm$^{-6}$ pc)          \df& 0.28 & 0.20 & $\pm0.10$ \nl
$kT_h$~(keV)\tablenotemark{c}  \df& 3.74 & 4.48 & $\pm0.90$ \nl
EM$_h$~(cm$^{-6}$ pc)          \df& 0.18 & 0.19 & $\pm0.02$ \nl
[Ne]\tablenotemark{e}          \df& 0.92 & 0.82 & $\pm0.40$ \nl
[O]                            \df& 0.28 & 0.39 & $\pm0.12$ \nl
[Fe]                           \df& 0.13 & 0.18 & $\pm0.08$ \nl
$\chi^2$~(d.o.f.)              \df& 327.4 (239) & 617.3 (403) & \nl
\enddata
\tablenotetext{a}{A third component was needed to fit the SNR spectrum,
as described in the text.}
\tablenotetext{b}{Parameter was fixed (Harnden \etal\ 1985).}
\tablenotetext{c}{The subscripts $l$ and $h$ refer to the low and high
                  temperature components respectively.}
\tablenotetext{d}{Unabsorbed incident flux in the 0.1--7.0~keV energy range,
                  in units of 10$^{-14}$ erg \pcm{2}\ \psec\ \parcmin.}
\tablenotetext{e}{Elemental abundances are given w.r.t. solar abundance.
                  Unlisted elements were fixed at solar abundance.}
\end{deluxetable}

\end{document}